\documentstyle[emulateapj,apjfonts,epsfig]{article}

\newcommand{\ergsec}{erg sec$^{-1}$}

\begin{document}

\title{Transient X-ray Binaries in Elliptical Galaxies}

\author{Anthony L. Piro}
\affil{Department of Physics, Broida Hall, University of California
	\\ Santa Barbara, CA 93106; piro@physics.ucsb.edu}
%\email{piro@physics.ucsb.edu}

\and

\author{Lars Bildsten}
\affil{Institute for Theoretical Physics and Department of Physics,
Kohn Hall, University of California
	\\ Santa Barbara, CA 93106; bildsten@itp.ucsb.edu}
%\email{bildsten@itp.ucsb.edu}

\begin{abstract}
 
{\it Chandra} observations of elliptical galaxies have revealed large
numbers of Low Mass X-ray Binaries (LMXBs) accreting at rates $>
10^{-9} M_\odot$ yr$^{-1}$. One scenario which generates this
$\dot{M}$ from an old stellar population is nuclear driven mass
transfer onto a neutron star or black hole from a Roche lobe filling
red giant. However, in our Galaxy, most of these systems accrete
sporadically as transients due to a thermal instability in the
accretion disk. Using the common criterion for disk instability
(including irradiation), we find that this mode of mass transfer
leads to transient accretion for at least $75\%$ of the binary's
life. Repeated {\it Chandra} observations of elliptical galaxies
should reveal this population. The recurrence times might be very long
\mbox{($\sim$ 100-10,000 years} depending on the orbital period at the
onset of mass transfer), and outbursts might last for 1-100
years. Mass transfering binaries can also be formed in old populations
via interactions in dense stellar environments, such as globular
clusters (GCs). These tend to have shorter orbital periods and are
more likely stable accretors, making them apparently a large fraction
of the elliptical's LMXB population.

\end{abstract}


\section{Introduction}

  {\it Chandra } observations of nearby elliptical and S0 galaxies
(e.g. Sarazin, Irwin \& Bregman 2000; Blanton, Sarazin \& Irwin 2000;
Sarazin, Irwin, \& Bregman 2001;
Finoguenov \& Jones 2001; Angelini et al. 2001; Kraft et
al. 2001) often find $\approx $ 20 X-ray sources with luminosities
greater than $10^{38}$ \ergsec and even more with luminosities greater
than $10^{37}$ \ergsec. For a neutron star (NS) accretor, this
corresponds to accretion rates as large as $10^{-8} M_\odot$
yr$^{-1}$, implying a short lifetime compared to the age of the
stellar population. One way to generate such rapid accretion rates in
an old stellar population (neglecting formation of tight binaries in
dense environments, such as globular clusters) is to have a Roche
lobe-filling binary driven by nuclear expansion of a low-mass star on
the red giant branch (orbital periods are days or longer). Ritter
(1999) has shown an analytic solution for such a Low Mass X-Ray Binary
(LMXB), which is further simplified in ellipticals where the initial
companion masses should all be $\approx 0.9M_\odot$ for an old
population. 

  However, such wide binaries are subject to thermal instabilities in
the accretion disk (King, Kolb \& Buderi 1996; see Dubus, Hameury \&
Lasota 2001 for an up-to-date discussion) that cause transient
accretion.  Using the current instability criterion, we follow the
evolution of such an LMXB and find that over 75\% of its life is spent
as a transient. We thus expect that a large fraction of the field
LMXBs in elliptical galaxies are long-term transients. Our ability to
detect them with repeated {\it Chandra} observations depends on their
outburst durations, which can be as long as years.

 We overview the binary evolution scenario in \S 2 and summarize the
challenge of finding the bright, blue optical counterparts to these
LMXBs in elliptical galaxies. In \S 3, we apply the current
understanding of accretion disk instabilities in X-ray sources and
estimate the outburst recurrence times. In \S 4, we compare our work
to galactic systems, noting that the recently discovered galactic
transients with outburst durations in excess of ten years
(e.g. Wijnands 2002) might well be wide binaries. We close in \S 5
with our conclusions and a discussion of alternative binary scenarios
that lead to more stable mass transfer, especially relevant for those
LMXBs found in globular clusters (Sarazin, Irwin \& Bregman 2001;
Angelini, Loewenstein \& Mushotzky 2001).

\section{Binary Evolution with Red Giant Donors}

  We only consider binaries with initial orbital periods $\ge$1
day. If the binary is tighter ($P_{\rm orb} < 1$ day), rapid
accretion may be initiated due to orbital angular momentum loss from
magnetic braking, resulting in mass transfer while a low-mass star is
still on the main sequence (see Tauris \& Savonije 1999). Whenever
there are multiple parameters to choose, we always take the value that
maximizes the stability of the accretion disk. This assures that
our calculated fraction of the evolutionary lifetime spent as a
transient is a firm lower limit. 

  We use Ritter's (1999) analytical solution for the LMXB evolution,
which assumes that the donor star is Roche lobe filling on the red
giant branch and expanding due to an increasing luminosity. The mass
transfer is slow enough that the donor star always remains close to
thermal equilibrium; thus $R_2 = R_{2,R}$, where $R_2$ is the radius
of the donor star and $R_{2,R}$ is the radius of its Roche lobe. For
simplicity, we assume that mass and angular momentum are conserved,
which implies that at the end of the LMXB's evolution the NS (of mass
$M_1$) will have a mass of 1.7-2.1$M_\odot$. We assume that the
initial donor mass is $0.9 M_\odot$ with a population I metallicity,
consistent with recent determinations of the age and metallicity of
the predominant stellar populations in these elliptical galaxies
(Trager et. al. 2000). 

  Once the evolution is solved, all parameters of the binary can be
found in terms of the orbital period at the onset of mass transfer,
$P_i$, which is measured in days, and the donor's mass, $M_2$, which
is measured in solar masses. The mass transfer rate within the binary
is
\begin{eqnarray}
        - \dot{M_2}=6.4 \times 10 ^{-10}
                M_\odot \textrm{ yr}^{-1}
                \mbox{ }P_i ^{ 14 / 15 }
                \left( \frac{M_2}{0.9} \right) ^{ - 7 / 3 }
                \nonumber \\
                \times \left( \frac{2.3 - M_2}{1.4} \right) ^{ - 14 / 5 }
                \left( \frac{5}{3} \frac{1}{M_2} - \frac{2}{2.3-M_2}
                         \right) ^{-1},
	\label{eqn_mdot}
\end{eqnarray}
where $1.4M_\odot$ is the initial NS mass and $2.3M_\odot$ is the
total mass of the binary. Since we will show in \S 3 that the disk is
likely unstable, we will not concern ourselves here with the implied
super Eddington accretion rates at large $P_i$ 
should this mass transfer rate proceed
onto the NS. Using equation ($\ref{eqn_mdot}$) the evolution of the
binary system is shown for varying $P_i$ by the
solid lines in Figure 1. The long dashed line is the boundary where
the hydrogen envelope of the donor has vanished, halting the
evolution at the ending orbital period noted.  The evolving orbital period is
\begin{eqnarray}
         P_{\rm orb} = P_i
                \left( \frac{M_2}{0.9} \right) ^{-3}
                \left( \frac{2.3 - M_2}{1.4} \right) ^{-3}, 
\end{eqnarray}
shown in Figure 2 by the solid lines. 

 For such an LMXB in an elliptical galaxy, we expect: (1) the orbit
must be wide in order to reach such high $\dot M$'s, and (2) the
reprocessing of the X-rays by the large accretion disk should make it
a bright, very blue object. At the Eddington luminosity, an X-ray
binary with $P_{\rm orb}=10$ days (much like Cygnus X-2) has $M_V=-2$
(van Paradijs \& McClintock 1994) and is blue ($B-V\approx 0,
U-B\approx -1$; van Paradijs \& McClintock 1995).  For NGC 4697 (at a
distance modulus of $\approx 30.35$; Tonry et al. 2001), this
corresponds to an apparent V magnitude of 28.4.
 
  Assuming that the LMXB is not associated with a globular cluster, the
brightest star which could confuse a search for the optical
counterpart would be a red giant branch star just about to reach
helium core ignition (a TRGB star). For metallicities $[{\rm
Fe/H}]<-1$, $M_V(TRGB)\approx -2.5$, whereas for $[{\rm Fe/H}]\approx
-0.7$, $M_V(TRGB)\approx -1.5$ (Saviane et al. 2000). For a slightly
metal-rich star (such as is likely the case for these ellipticals;
Trager et al. 2000), the peak brightness is even fainter,
$M_V(TRGB)\approx -1.0$ (Garnavich et al. 1994). Hence, the optical
counterpart to a wide LMXB will be brighter at $V$ (and especially so
in $B$ or $U$) than any red giant branch star in an elliptical galaxy.

 The surface brightness of an elliptical galaxy would make such a
search difficult in the central regions. However, beyond $\approx$ 2
arcminutes, the surface brightness at $V$ (in NGC 4697, for example) is
below the sky level (for HST) of 22-23 ${\rm Mag}/{\rm arc-sec^2}$
(Goudfrooij et al. 1984). In that limit, and using the measured {\it
Chandra} position of sources at such locations, such a wide (10 day)
binary would be detectable at B or V in a $5\times 10^4\ {\rm s}$
observation with the {\it Advanced Camera for Surveys} on HST.  Of
course, such optical counterparts are easy to find in the nearby bulge
of M31, where $V\approx 22$.

\section{Disk Instability, Transient Fraction, and Recurrence Times}

  If the disk is thermally stable and transfers matter to the NS at the
supplied rate, there is nearly a one-to-one relation between $P_i$ 
and the X-ray luminosity (see Figure 1) that would allow the 
orbital period distribution to be inferred from the luminosity
distribution (Webbink et al. 1983). However, this argument fails
because these wide binaries are almost always transient accretors. 

  The usual condition used for the disk instability is that some portion
of the disk is below the hydrogen ionization temperature ($\approx
6500$ K). Since the effective temperature of the disk decreases with
increasing radius, it is usually adequate to check that the outer
radius of the disk is below the critical temperature. In X-ray
binaries, however, this temperature is fixed by irradiation 
from the central accreting source (van Paradijs 1996). The
surface temperature is then estimated as (Dubus et al. 2001) 
\begin{eqnarray}
        T_{\rm irr}^4 \approx
		\frac{C \eta \dot{M_1} c^2 }{ 4 \pi \sigma R^2 }, 
\end{eqnarray}
where $R$ is the disk radius, $\sigma$ is the Stefan-Boltzmann
constant, $\eta=0.1$ is the conversion efficiency for accretion onto
the NS, and $C$ measures the fraction of the X-ray luminosity that
reaches the disk (and contains information on the irradiation
geometry). We take $C=5.7\times 10^{-3}$ (slightly larger than
the preferred value, $C=5\times 10^{-3}$ of Dubus et al. 2001) so as
to maximize the persistent fraction. The disk is assumed to extend out
to 70\% of the Roche lobe around the NS (King et al. 1997).

  Setting $T_{\rm irr}=6500 $ K allows us to solve for the critical time,
accretion rate, and orbital period at which the instability sets in as
a function of the donor's mass.  The critical mass transfer rate is then
\begin{eqnarray}
        - \dot{M}_{\rm 2,crit}  = 3.3 \times 10 ^{-9}
		M_\odot \textrm{ yr}^{-1}
		M_2 ^{ 14 / 9 }
                % \nonumber \\
                \left( 2.3 - M_2 \right) ^{ - 14 / 9 }
		\nonumber \\
                \times \left( \frac{5}{3} \frac{1}{M_2} - \frac{2}{2.3-M_2}
                         \right) ^{-10/3}, 
\end{eqnarray}
which is used for the instability boundary shown in Figures 1 and 2 as
the dotted line. The critical orbital period is 
\begin{eqnarray}
         P_{\rm orb,crit} = 5.5 \textrm{ days }
                M_2^{ 7 / 6 }
                \left( 2.3 - M_2 \right) ^{ - 5 / 3}
		\nonumber \\
		\times \left( \frac{5}{3} \frac{1}{M_2}
			- \frac{2}{2.3-M_2} \right) ^{ -5 / 2 },
\end{eqnarray}
and this is used for the instability boundary in Figure 2.  The
fraction of time spent transient is then found (numerically to better
than a percent) as a
function of the initial orbital period,
\begin{eqnarray}
 	{\rm Transient \ Fraction} = -0.0500 \chi ^2 + 0.2385 \chi + 0.7655,
\end{eqnarray}
where $\chi = \log_{10} \left( P_i / {\rm d} \right)$.  For $P_i$ 
in excess of a day, the binary is transient at least $75\%$ of
the time, and this does not include binaries containing black holes
which are transient almost all the time (e.g. King et al. 1996). 

  We estimate a maximum recurrence time for outbursts by assuming the
outside-in model. We assume that the binary is quiescent until a
critical surface density in the outer accretion disk (at radius $R$)
has been reached (Dubus, Hameury,\&  Lasota 2001),
\begin{eqnarray}
        \Sigma_{\rm max} = 644 \textrm{ g cm}^{-2}
                ( M_1/M_\odot )^{ - 0.37 }
                ( R/R_\odot )^{ 1.11 },
\end{eqnarray}
where we assume that the viscosity parameter $\alpha=0.1$ and that
there is no irradiation of the disk in quiescence. This expression 
gives a maximum recurrence time
\begin{eqnarray}
        t_{\rm rec,max} =  15 \textrm{ yrs }
                P_i ^{1.14}
                M_2 ^{-3.88}
                \left( 2.3 - M_2 \right) ^{-2.75}
		\nonumber \\
		\times
                \left( \frac{5}{3} \frac{1}{M_2}
                        - \frac{2}{2.3-M_2} \right), 
\end{eqnarray}
that should be viewed with some caution given the uncertainties in the
accretion disk physics. 
This recurrence time is $\sim$ 100-10,000 years (depending on the initial
orbital period; see Figure 2) and is consistent with current estimates
from observations of galactic ms radio pulsar/He white dwarf binaries
(Ritter \& King 2002). Our estimate is maximal because it is possible
(and maybe even likely) that the LMXB may outburst before the outer disk
reaches the critical mass, such as an outburst initiated in the inner disk.  

\section{Comparison with Galactic Sources} 

  One difficulty with comparing these evolutionary models with
observations is the lack of long orbital period LMXBs in our galaxy.
Many persistent sources have low enough orbital periods ($\sim1-20$
hours) that they are probably not accurately described by the
evolutionary model used above. However, a lack of persistent sources
at $P_{\rm orb} > 1 $ day is in agreement with our calculations and
generally expected from previous work (for example King, Frank, Kolb \&
Ritter 1997).  Figure 1 shows that such NS binaries spend most of
their lifetime as transient rather than persistent sources ($>75\%$ of
the time).  Though all these relations are consistent, it is still
hard to claim that the instability criterion has been fully
scrutinized by the galactic population.

  Indeed, the outburst durations of these systems might be so long that
they appear to us as ``persistent'' sources (with accretion rates most
likely in excess of the mass transfer rate expected from binary
evolution) until they suddenly disappear. Indeed, one such source (and
maybe others; Wijnands 2002) has now been clearly identified:
KS~1731-260 (Wijnands et al. 2001; Rutledge et al. 2002). It was
accreting at $\dot M\approx 3\times 10^{-9}M_\odot {\rm yr^{-1}}$ for
at least a decade prior to its sudden decline in luminosity by a
factor of $>10^4$ (Rutledge et al. 2002). Analysis of the immediate
quiescent emission in terms of thermal emission from the NS surface
yielded constraints on the recurrence time from 300-1,000 years
(Wijnands et al. 2001; Rutledge et al. 2002). Until this system's
orbital period is found (current knowledge of the infrared counterpart
points to a possibly evolved companion; Orosz, Bailyn \& Whitman 2001)
we will not be able to say whether it is an LMXB of the type we are
discussing. However, its existence does point to a likelihood of
finding such similar systems in elliptical galaxies.

  One galactic binary that does have a long orbital period is GRO
J1744-28 (11.8 days; Finger et al. 1996).  This system is transient
and was shown by Bildsten and Brown (1996) to have a companion with
mass $M_2 \approx 0.334 M_\odot$ and core mass \mbox{$M_c \approx 0.22
M_\odot$} (if the companion is a population I metallicity star). When
this binary's evolutionary path is predicted and plotted, it is found
to be transient (as it is) with a maximum outburst recurrence time of
\mbox{$\sim$1,000 yrs} (see dot-dashed line in Figure 2), consistent
with a single outburst.

\section{Conclusions and Discussion}

  We have shown that the expected mass transfer scenario for low mass
X-ray binaries in the field of an elliptical galaxy (where the donors
must be $<M_\odot$) that yield $\dot M\approx 10^{-8} M_\odot
{\rm yr^{-1}}$ (as observed by {\it Chandra}) have thermally unstable
disks. Such binaries are expected to be transient accretors with
maximum recurrence times of 100-10,000 years for at least $75\%$ of
their lifetimes. When active, it appears likely that they would easily
reach the NS's Eddington limit.

  Our inability to calculate the duration of the bright outbursting
phase makes it difficult to predict how many such transients will be
identified in repeated {\it Chandra} visits. If the duty cycle is 
bright for ten years out of a thousand (like implied for KS 1731-260;
Rutledge et al. 2002), and there are 100 such binaries, then a year
later, roughly 10 would have faded into quiescence to be replaced by
new transients. This is not that far off from what was seen over a
five month interval in Cen A (Kraft et al. 2001), where four sources
(out of 120 exposed in both epochs) were either non-detectable
($L_x<(1-2)\times 10^{36} \ {\rm erg \ s^{-1}}$) or appeared bright
($L_x>4\times 10^{37} {\rm erg \ s^{-1}}$) in either epoch. There were a
larger number ($\approx 35$) which showed significant variability
(Kraft et al. 2001).

  Such comparisons are also complicated by the discovery that many of
the X-ray sources in ellipticals are coincident with GCs (Sarazin et
al. 2001; Angelini et al. 2001; White, Sarazin \& Kulkarni 2002). 
Other possible accretion scenarios then include accretion from
lower-mass main sequence stars that are formed in this dense stellar
environment (see Rasio, Pfahl \& Rappaport 2000 for a recent
discussion). These tight systems would most likely be persistent
accretors, thus apparently reducing the fraction of transients amongst
the detected sources. This might just be an impression however, as if
the duty cycle of the field transients is really 1\%, then the
underlying population is indeed larger than that in GCs. Namely, it is
possible that the GC population is only {\it apparently } a large
fraction because LMXBs in GCs are persistently bright. The census of
the much more numerous population of field transients can only be
taken over a few thousand years.

  We thank G. Dubus, J. Irwin, H. Ritter, and C. Sarazin for helpful
discussions and the anonymous referee for a very informative critique of
our work. This research was supported by NASA via grant NAG 5-8658
and by the NSF under Grants PHY99-07949 and AST01-96422. L. B. is a
Cottrell Scholar of the Research Corporation.

\begin{figure}
\plotone{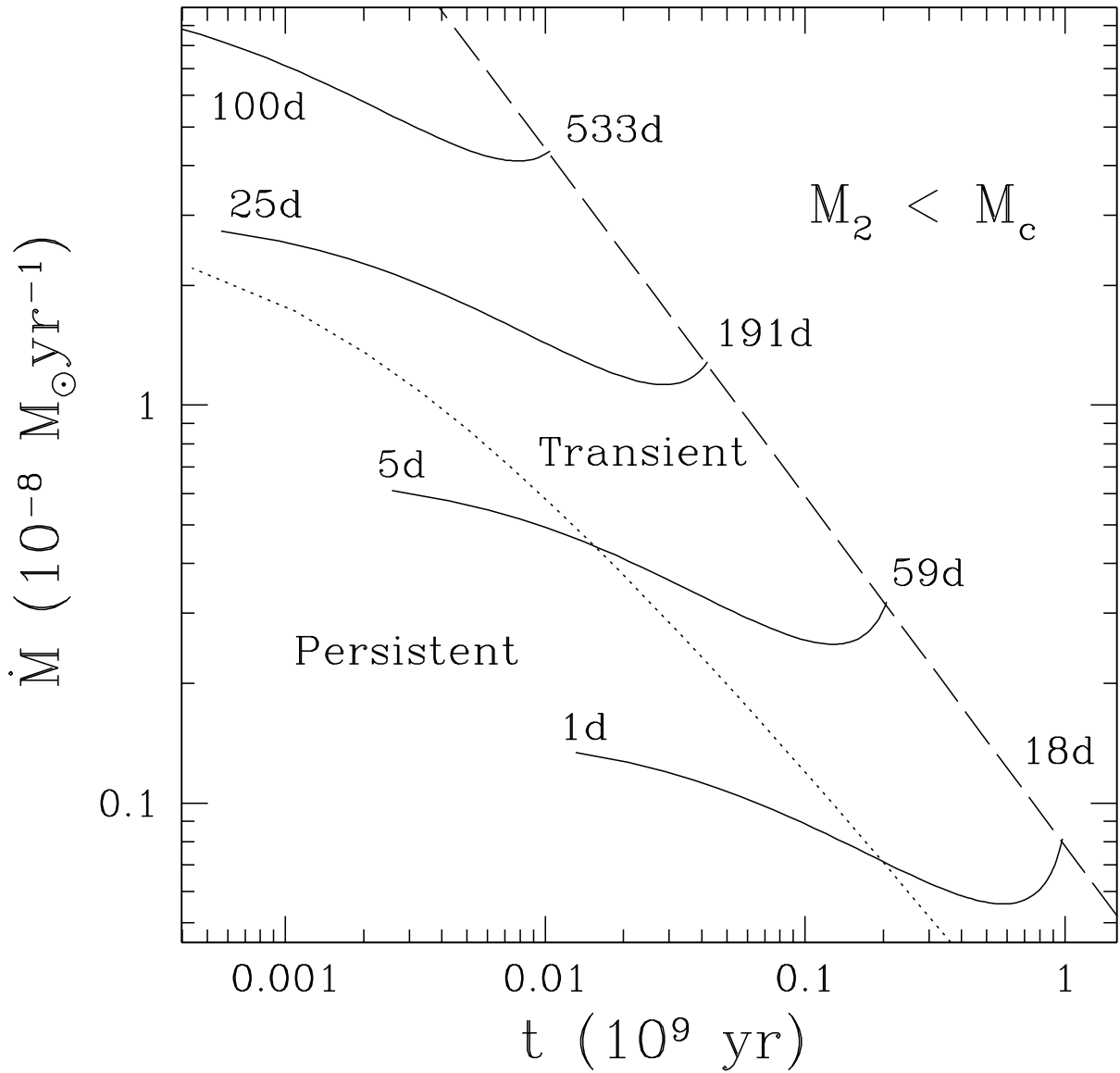} 
\figcaption{Mass transfer rate as a function of
time for LMXBs with red giant donors. The solid lines are the
evolutionary path of the binary given an initial orbital period when
Roche lobe overfilling begins (evolution is from left to right).
The dashed line is the boundary where the Hydrogen envelope of the
donor has vanished, thus halting the evolution. 
The dotted line is the boundary at which the
accretion disk instability begins.}
% \label{fig1}
\end{figure}

\begin{figure}
\plotone{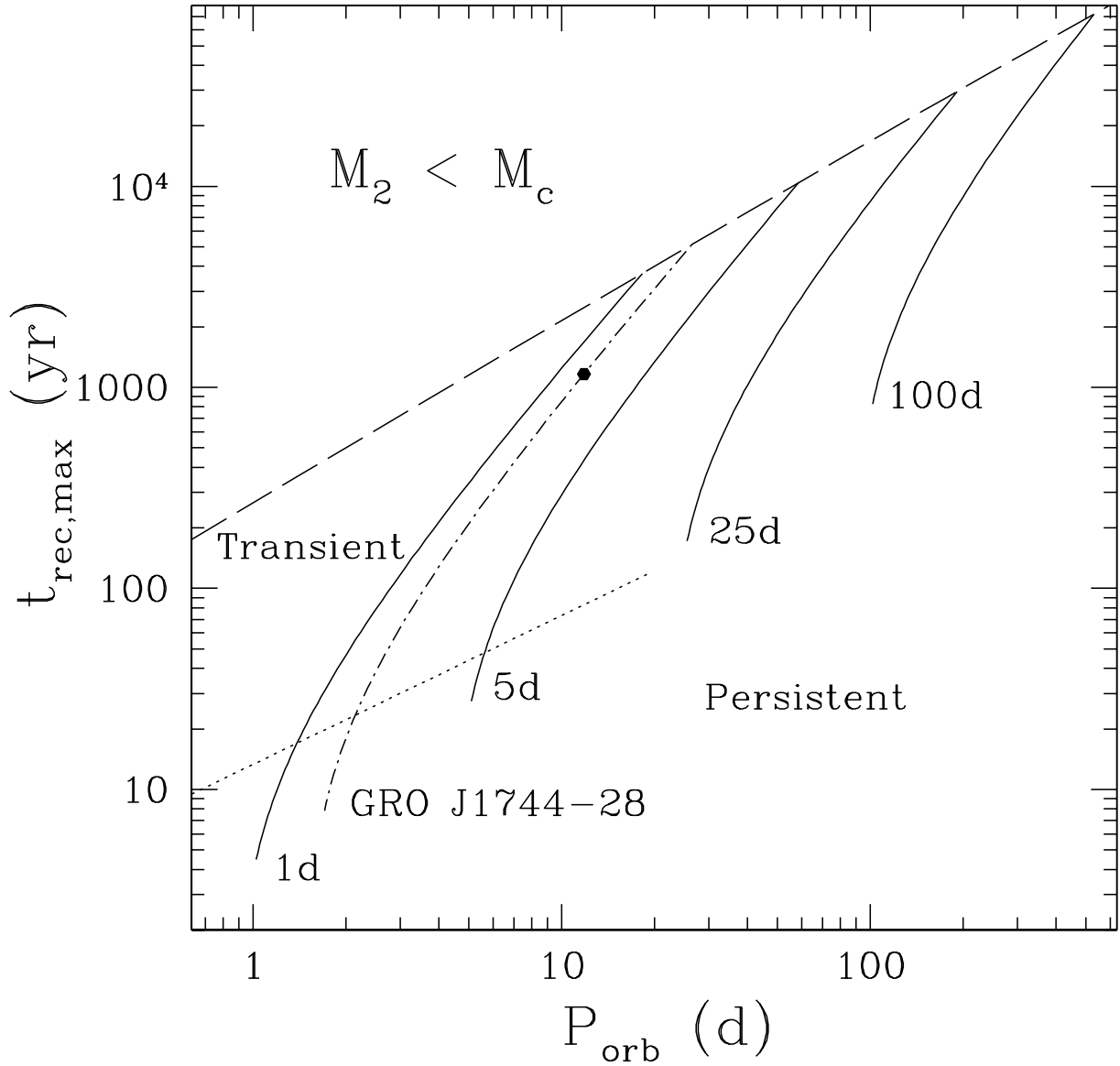}
\figcaption{Estimated 
maximum recurrence time as a function of the orbital period.
The curves are the same as in Figure 1. The additional dot-dashed line is
the predicted evolutionary path of GRO J1744-28. The filled circle
denotes this binary's current orbital period of 11.8 days.}
% \label{fig2}
\end{figure} 

\end{document}